\documentclass[floatfix,aps,showpacs,twocolumn,pra]{revtex4}
\usepackage{graphicx}
\usepackage{amsmath}
\usepackage{amssymb}

\begin{document}

\newcommand{\x}{{\bf x}}
\renewcommand{\k}{{\bf k}}
\def\mn#1{\marginpar{*\footnotesize \footnotesize #1}{*}}
\def\isum{{\sum\!\!\!\!\!\!\!\!\int}}
\def\bfzero{{\bf 0}}
\newcommand{\be}{\begin{equation}}
\newcommand{\ee}{\end{equation}}
\newcommand{\bea}{\begin{eqnarray}}
\newcommand{\eea}{\end{eqnarray}}
\newcommand{\bdm}{\begin{displaymath}}
\newcommand{\edm}{\end{displaymath}}
\newcommand{\bfk}{{\bf k}}
\def\hF{{\bar F}}
\newcommand{\bfp}{{\bf p}}
\newcommand{\ha}{{\hat a}}
\newcommand{\bfr}{{\bf x}}
\newcommand{\bfz}{{\bf 0}}
\newcommand{\tG}{{\widetilde G}}
\newcommand{\veps}{\varepsilon}
\newcommand{\non}{\nonumber \\}
\newcommand{\nsh}{{\delta}}
\newcommand{\bfx}{{\bf x}}
\newcommand{\nln}{\nonumber\\}
\def\pder#1#2{\frac{\partial #1}{\partial #2}}
\def\ep#1{\langle #1\rangle}
\def\lsim{\raise0.3ex\hbox{$<$\kern-0.75em\raise-1.1ex\hbox{$\sim$}}}

\title{Condensation temperature of interacting Bose gases with and without disorder}
\author{O.~Zobay}
\affiliation{Institut f\"ur Angewandte Physik, Technische Universit\"at
Darmstadt, 64289 Darmstadt, Germany}
\begin{abstract}
The momentum-shell renormalization group (RG) is used to study the condensation of interacting Bose gases without and with disorder. First of all, for the homogeneous disorder-free Bose gas the interaction-induced shifts in the critical temperature and chemical potential are determined up to second order in the scattering length. The approach does not make use of dimensional reduction and is thus independent of previous derivations. Secondly, the RG is used together with the replica method to study the interacting Bose gas with delta-correlated disorder. The flow equations are derived and found to reduce, in the high-temperature limit, to the RG equations of the classical Landau-Ginzburg model with random-exchange defects. The random fixed point is used to calculate the condensation temperature under the combined influence of particle interactions and disorder.
\end{abstract}
\maketitle

\flushbottom

The realization of Bose-Einstein condensation in ultracold atomic gases has renewed the interest in the quantum-statistical properties of interacting Bose gases around the phase transition. One particular focus of theoretical work in this area concerns the calculation of the critical temperature $T_c$ of the homogeneous Bose gas as a function of the scattering length $a$. Due to its nonperturbative character, this apparently simple problem turned out to present a considerable theoretical challenge. For a long time, the leading-order functional dependence of the shift in $T_c$ on the scattering length remained highly controversial. It was even unclear whether the interactions lead to an upward or downward shift in the critical temperature \cite{And04}. Only recently, a generally accepted result has emerged \cite{BayBlaHol99,HolBayBla01,ArnMooTom01}. It was shown that the critical temperature of an interacting Bose gas with spatial density $n$ obeys the relation
\bea\label{shift_Tc}
\frac{T_c}{T_c^0}&=&1+c_1 an^{1/3}+[c_2'\ln (an^{1/3})+c_2''] a^2n^{2/3}\non
&& +O(a^3n),
\eea
where $T_c^0$ denotes the critical temperature of an ideal Bose gas with the same density. To calculate the coefficients $c_i$, one can make use of dimensional reduction \cite{And04,Zin89} and perturbatively match the full quantum-mechanical problem to a classical field theory with a lower-- (i.e., three-) dimensional action \cite{ArnMooTom01}. The coefficient $c_2'$ can then be determined by perturbation theory alone, whereas the evaluation of $c_1$ and $c_2''$ requires nonperturbative input from the classical theory. This input was calculated in Refs.\ \cite{ArnMoo01,ArnMoo01b,KasProSvi01} with so far unsurpassed accuracy by means of Monte Carlo lattice simulations.

Even after the qualitative and quantitative behavior of the $T_c$ shift had been solidly established \cite{BayBlaHol99,HolBayBla01,ArnMooTom01,ArnMoo01,KasProSvi01}, a large number of papers continued to appear that presented further calculations of critical properties of interacting Bose gases. The main purpose of these works was not to correct or improve the previous results, but to introduce, test, and refine alternative approaches. The applied methods comprise, e.g., linear $\delta$ expansion \cite{BraRad02b,BraRad02,SouPinRam01,SouPinRam02,KnePinRam03,KnePin05}, variational perturbation theory \cite{Kle03,Kas03,Kas04}, and the exact renormalization group \cite{LedHasKop04,HasLedKop04,BlaGalWsc04}.

The present paper follows this research line of investigating alternative approaches to studying the interacting Bose gas at criticality. One of its main motivations is the observation that most of the methods that have been used so far, e.g., the perturbative matching procedure of Ref.\ \cite{ArnMooTom01} for finding the second-order shifts as well as the nonperturbative calculations within the classical theory, are relatively involved. This raises the question of whether there exist alternative approaches that, on the one hand, are more straightforward to apply, but, on the other hand, still yield sufficiently accurate quantitative results and are also readily applicable to more complicated problems. In the present paper it is shown that the one-loop momentum-shell renormalization group (RG) method \cite{WilKog74} provides such an approach. To substantiate this claim, we will investigate two nontrivial problems: the second-order shifts in the critical temperature and chemical potential of the homogeneous Bose gas, and the condensation temperature of the Bose gas with delta-correlated disorder \cite{HuaMen92,LopVin02}.

The purpose of these studies is fourfold. (i) As already mentioned above, we would like to demonstrate that this RG method provides a transparent, elegant and efficient way to investigate critical thermodynamic properties of interacting Bose gases. First of all, quantitative results for, e.g., the critical temperature and chemical potential are readily obtained by numerically solving simple RG flow equations. The accuracy of the results is comparable to the other recent approaches mentioned above. Furthermore, the approach also allows one to obtain some qualitative insights into the essential behavior of critical quantities. To illustrate this point, it will be shown how the nonanalytic dependence of the critical chemical potential on the scattering length can be extracted analytically from the flow equations. 

(ii) We present the first second-order calculation of the critical temperature and chemical potential that does not rely on perturbative matching to the three-dimensional $O(2)$ scalar field theory. Most calculations of $c_1$ and, to the author's knowledge, all previous evaluations of $c_2''$ \cite{ArnMoo01b,SouPinRam02,KnePinRam03,Kas04,KnePin05} proceed by determining nonperturbative parameters of the classical three-dimensional (3D) theory and relating them to the quantum-mechanical problem through the matching procedure of Refs.\ \cite{ArnMooTom01,ArnMoo01,ArnTom01}. It would thus be desirable to evaluate the critical temperature shift in a way which is independent of this procedure. The present approach provides such an independent derivation since it works with the full $3+1$-dimensional action and does not make use of dimensional reduction.

(iii) By investigating the Bose gas with delta-correlated disorder \cite{HuaMen92,LopVin02}, we show that the RG approach can easily be extended to the study of more complicated systems. This model problem has been of interest recently in order to understand qualitative features of certain systems in random environments, such as He4 in porous media. The RG flow equations are derived with the help of the replica method. It is shown that in the high-temperature limit they reduce to the flow equations describing classical spin systems with random exchange disorder \cite{HarLub74,Lub75,GriLut76,Khm76,Her85}.

(iv) Using the random fixed point of the flow equations, we calculate the condensation temperature in the presence of particle interactions and disorder. The only determination of $T_c$ so far \cite{LopVin02} is of perturbative nature and neglects the atomic interactions. However, it is known that the interactions are an essential part of the model since they stabilize the disordered Bose gas against collapse \cite{HuaMen92}. They also play a crucial role in the critical flow obtained from the RG equations. It would thus be more satisfactory to calculate the condensation temperature within an approach that directly includes the interactions. This is accomplished by our RG method.

The approach presented in this paper is based on the application of the momentum-shell RG method \cite{WilKog74} to a weakly interacting boson system. Since the 1970s, a number of studies have considered this question (see, e.g., Refs.\  \cite{Sin75,CreWie83,FisHoh88,KolStr92,BijSto96,AndStr99,Alb01}). Whereas some of these works mainly focussed on the critical exponents \cite{Sin75,CreWie83,AndStr99}, others used the RG equations to investigate critical thermodynamic quantities. In Ref.\  \cite{FisHoh88}, the critical temperature of a two-dimensional (2D) Bose gas was examined, while Refs.\ \cite{BijSto96} and \cite{Alb01} studied critical properties of the 3D gas using the RG equations for the {\it condensed} phase. These equations, however, do not yield the correct behavior for, e.g., the critical temperature \cite{Haq03}. After clarifying the relation between the flow equations for the condensed and the uncondensed phase \cite{MetZobAlb03}, the latter were successfully used to investigate the critical temperature of trapped Bose gases \cite{MetZobAlb04,ZobMetAlb04}. 

The particular appeal of the RG approach lies in its conceptual simplicity combined with the fact that the thermodynamic properties of interest are easily extracted numerically from the flow equations. As discussed in detail in Refs.\ \cite{MetZobAlb04} and \cite{MetAlb02}, the approach has some shortcomings: it is only approximate, and some of the underlying assumptions and approximations are not fully controllable and difficult to improve systematically. Nevertheless, these problems are outweighed by the advantages of the method.

We first consider the ordered homogeneous Bose gas. Its partition function can be written as a functional integral
\begin{equation}
Z = \int {\cal D}\phi^*\int {\cal D}\phi\,  e^{- \frac{1}{\hbar}  S[\phi,\phi^{*}]},
\label{partfunc}
\end{equation}
with the Euclidean action
\begin{eqnarray}
S[\phi, \phi^*] &=&
 \int_0^{\hbar \beta }\! d\tau\!\int_V \! d^3 {\bf x}
\left\{ \phi^* (\tau, {\bf x})
\left[\hbar\frac{\partial}{\partial \tau} - \frac{\hbar^2}{2m} \nabla^{2} \right.\right.\nonumber \\
&& \left.    - \mu\Big]
 \phi(\tau, {\bf x}) + \frac{g}{2} |\phi(\tau, {\bf x})|^4
 \right\}{\ },
\label{action}
 \end{eqnarray}
$\mu$ denoting the chemical potential, $\beta$ the inverse temperature, $m$ the particle mass, and $g$ the coupling constant for the $s$-wave contact interaction. The bosonic fields $\phi(\x,\tau)$ are periodic in $\tau$ with period $\hbar\beta$ and obey spatial periodic boundary conditions.
We Fourier-decompose the bosonic fields according to
\be
\phi(\x,\tau)=\frac 1{(\hbar\beta V)^{1/2}} \sum_{\k,n}a_{\k,n}\exp[i(\k\cdot\x -\omega_n \tau)],
\ee
with $\omega_n=2\pi n/\hbar\beta$ and the cutoff condition $|\k|\le \Lambda$, and split them into a slow and a fast field component
$
\phi(\x,\tau)=\phi^<(\x,\tau)+\phi^>(\x,\tau),
$
where the fast field $\phi^>$ only contains momenta within a thin shell $\Lambda-d\Lambda < |\k| < \Lambda$ near the cutoff. The Euclidean action is then expanded up to second order in $\phi^>$. Carrying out the functional integration over $\phi^>$, expanding the result up to second order in the coupling strength, and using lowest-order derivative expansion, one can write the partition function in the original form (\ref{partfunc}), (\ref{action}), where the integration is now over the slow field only and the parameters $\mu$ and $g$ have received an (infinitesimal) renormalization. Note that this approach retains the contribution of all Matsubara frequencies and thus the full quantum-mechanical character of the problem. After rescaling all quantities to restore the original cutoff, the procedure is repeated until the bosonic fields are integrated out completely. The concomitant change of $\mu$ and $g$ is described by the flow equations
\bea
\label{flow1}\frac{dM}{dl}&=&2M- 2\frac{\tG}{2\pi^2} bN_{BE},\\
\label{flowtG}\frac{d\tG}{dl}&=& \tG-\frac{\tG^2b}{2\pi^2}\left[4bN_{BE}(N_{BE}+1)  +\frac{1+2N_{BE}}{2(E_>-M)}\right],
\eea
for the scaled quantities $M=\beta_\Lambda\mu$, $\tilde G=\Lambda^3\beta_\Lambda^2 g/\beta$, $b=\beta/\beta_\Lambda$ with the inverse cutoff temperature $\beta_\Lambda=m/\hbar^2\Lambda^2$, the scaled Bose-Einstein distribution $N_{BE}=\{\exp[b(E_>-M)]-1\}^{-1}$ and the scaled cutoff energy $E_>=1/2$. The flow parameter $l$ describes the reduction of the original cutoff according to $\Lambda e^{-l}$. Details of the derivation of Eqs.\ (\ref{flow1}) and (\ref{flowtG}) and further discussions can be found in Refs.\ \cite{BijSto96,MetZobAlb03,MetZobAlb04,MetAlb02}. Similar equations were also used, e.g., in Refs.\ \cite{Sin75,CreWie83,FisHoh88,KolStr92}.

The flow equations (\ref{flow1}) and (\ref{flowtG}) possess a nontrivial critical point at $(M^*,\tG^*)=(1/12,5\pi^2/72)$. Flow trajectories asymptotically reaching the fixed point represent a system at the phase transition, and their initial values $[M(0),\tG(0),b(0)]$ correspond to critical thermodynamic parameters. The initial value $\tG(0)$ is related to the $s$-wave scattering length via $a \Lambda = \tilde G(0)b(0)/[4\pi + 2\tilde G(0) b(0)/\pi]$ \cite{BijSto96,MetZobAlb04}. 

Using this framework, we now investigate the critical temperature in the limit of small $\ha=a/\lambda_T$ with $\lambda_T=(2\pi\hbar^2\beta/m)^{1/2}$ the thermal wavelength. Instead of calculating $T_c$ directly, we initially work out the shift $\nsh=(n\lambda_T^3)_c -\zeta(3/2)$ in the scaled critical density. This quantity is determined by first of all calculating the thermodynamic potential
\begin{equation}\label{Whom}
W_{\rm hom} = \frac{1}{2\pi^2 b(0)}\int_0^\infty dl e^{-3l}
\ln \left[1 - e^{- b(l) [E_> - M(l)] } \right] 
\end{equation}
for a critical trajectory, taking the derivative $s_{\rm hom}=-\partial{W_{\rm hom}}/\partial M(0)|_{b(0), {\tilde G}(0)}$ and finally evaluating $\nsh=[2\pi b(0)]^{3/2}s_{\rm hom}-\zeta(3/2)$ \cite{MetZobAlb04}. The behavior of the critical temperature is then obtained through an inversion procedure \cite{ArnMooTom01}. In addition, we will examine another nontrivial critical quantity, i.e., the (scaled) chemical potential $v\equiv (\beta\mu)_c = M(0)b(0)$ which was previously studied in Ref.\ \cite{ArnTom01}. 

It is interesting to note that in Ref.\ \cite{FisHoh88}, the (superfluid) transition temperature of a two-dimensional interacting Bose gas was calculated by a related RG method. This calculation was also based on the evaluation of the critical density as the derivative of the free energy. Some analytic arguments using flow equations similar to Eqs.\ (\ref{flow1}) and (\ref{flowtG}) were applied to work out the derivative and in this way to finally arrive at a qualitative estimate for $T_c$. In the present case, the flow equations (\ref{flow1}) and (\ref{flowtG}) can also be used to obtain some qualitative insights into the essential behavior of critical quantities. As discussed in the Appendix, one can easily show that they imply a functional form
\bdm
v(\ha) \simeq b_1^{(v)}\ha+ b_2^{(v)} \ha^2 \ln\ha + b_3^{(v)} \ha^2
\edm
for the dependence of the critical chemical potential on the scattering length.
This is in agreement with the result of Ref.\ \cite{ArnTom01}.

Our ultimate goal, however, is to perform a quantitative calculation of critical properties. Determining $\nsh$ and $v$ analytically through Eqs.\ (\ref{flow1})--(\ref{Whom}) is rather difficult since one would have to very accurately relate initial conditions to the asymptotic behavior across a highly nonlinear regime of propagation. We thus perform our studies via a numerical analysis. Careful numerical evaluation allows us to determine $\nsh(\ha)$ and $v(\ha)$ for values of $\ha$ less than $10^{-6}$. The data obtained in this way is then used to determine the functional behavior of $\nsh$ and $v$ up to second order in $\ha$ through a fitting procedure. The fitting is somewhat complicated by the fact that the individual numerical errors for $\nsh$ and $v$ do not appear to be independently Gaussian distributed so that a standard $\chi^2$ analysis cannot be performed. We thus adopt the procedure described in the following. 

We assume a certain functional form for $v$ and $\nsh$, e.g.,
\begin{subequations}\label{form}
\bea
\nsh(\ha) &=& b_1^{(\nsh)}\ha+ b_2^{(\nsh)} \ha^2 \ln\ha + b_3^{(\nsh)} \ha^2,\\
v(\ha) &=& b_1^{(v)}\ha+ b_2^{(v)} \ha^2 \ln\ha + b_3^{(v)} \ha^2,
\eea
\end{subequations}
and perform a least-square-fit using all available data points below an upper limit $\ha_0$. Carrying out the fit for different values of $\ha_0$, we obtain coefficient functions $b_i^{(\nsh,v)}(\ha_0)$. This approach allows us to deal with a number of issues.

(i) The actual functional form of $v(\ha)$ and $\nsh(\ha)$ also contains terms of higher order in $\ha$ that we are neglecting. For this reason, the functions $b_i(\ha_0)$ become dependent on $\ha_0$. Taking the limit $\ha_0\to 0$, however, the higher-order terms get less important, and we can read off the final values for $b_i$ as $\lim_{\ha_0\to 0} b_i(\ha_0)$. Furthermore, we find that the inclusion of higher-order terms in the fitting scheme can significantly reduce the $\ha_0$ dependence of the functions $b_i$, whereas the limiting values hardly change. However, we did not attempt to systemically investigate such higher-order terms.

(ii) The numerical data cannot be obtained with arbitrary precision but contain errors. This puts limitations on the practical feasibility of the procedure described in (i). For $v(\ha)$ this issue is not severe, since the initial conditions $M(0)$ and $b(0)$ can easily be calculated to very high accuracy by means of a bisection routine \cite{MetZobAlb04}. For $\nsh(\ha)$, however, one has to evaluate the derivative of the thermodynamic potential (\ref{Whom}) which adds a significant amount of numerical noise. The noise shows up in an erratic behavior of the functions $b_i^{(\nsh)}(\ha_0)$ for small arguments. Only for larger $\ha_0$, the functions become sufficiently smooth. We read off the limiting values $\lim_{\ha_0\to 0} b_i(\ha_0)$ from the smooth part, but have to allow for significantly larger errors bars of the coefficients $b_i$ for $\nsh(\ha)$ as compared to $v(\ha)$, especially for the second-order terms. To some extent, the problem can be alleviated by using very large data sets. Error estimates for the coefficients $b_i^{(\nsh,v)}$ are obtained from inspecting graphs of the functions $b_i(\ha_0)$.

(iii) The functional form of $v(\ha)$ and $\nsh(\ha)$ cannot be assumed to be known {\it a priori} [although the results of the Appendix suggest relations like (\ref{form})], and from the numerical analysis alone, one is not able to prove unambiguously that a certain form is indeed the correct one. Nevertheless, a useful fitting procedure should help to distinguish between more and less plausible behaviors. To show that our method is indeed able to this, we note that, for example, a slight change in the fit function from $\ha^2 \ln\ha$ to $\ha^{1.99} \ln\ha$ has a strongly adverse effect on the behavior of the coefficients $b_i^{(v)}(\ha_0)$ for $v$, i.e., they no longer smoothly reach a limiting value for $\ha_0\to 0$. By performing similar tests with varying functional forms, we have verified that the behavior indicated above indeed provides a very good representation of the data. Finally, we also mention that the fitting procedure was successfully tested with sets of simulated noisy data.

The results of the numerical analysis can be summarized as follows. We find that the behavior of $v(\ha)$ and $\nsh(\ha)$ for small $\ha$, as obtained from our RG approach, is well compatible with a functional form of the kind (\ref{form}), whereas other suitable forms could not be identified. For the coefficients we find the values
\bdm
\begin{array}{rclll}
b_1^{(\nsh)}&=&-5.303&(\pm 0.005)&[-7.1],\\
b_2^{(\nsh)}&=&-177&(\pm 5)&[-146.81],\\
b_3^{(\nsh)}&=&-720&(\pm 20)&[-587],\\
b_1^{(v)}&=&10.4495&(\pm 0.0003)&[10.4495],\\
b_2^{(v)}&=&122.2&(\pm 0.2)&[100.53],\\
b_3^{(v)}&=&425.5&(\pm 0.7)&[386.71],
\end{array}
\edm
with the numbers in square brackets giving the values of Refs.\ \cite{ArnMooTom01} and \cite{ArnTom01}.
Using the inversion formulas of Ref.\ \cite{ArnMooTom01}, the coefficients of Eq.\ (\ref{shift_Tc}) are obtained as
\bdm
\begin{array}{rclll}
c_1&=&0.9826&(\pm 0.001)&[1.32],\\
c_2'&=&23.8&(\pm 0.7)&[19.75],\\
c_2''&=&91&(\pm 3)&[75.7].
\end{array}
\edm
We find that the deviations of RG values from the results of Refs.\ \cite{ArnMooTom01} and \cite{ArnTom01} range from 10\% to 26\% with the largest differences occurring for the coefficients $b_1^{(\nsh)}$ and $c_1$. The accuracy of these results is thus comparable to some of the other recent approaches mentioned above. Earlier approaches even had deviations of far more than 100\% from the current best values \cite{And04}. Furthermore, it should be emphasized again that our calculation of the second-order shifts is completely independent from previous works such as Ref.\ \cite{ArnMooTom01} and does not make use of dimensional reduction. It also appears to be considerably simpler than these approaches.

We now turn to the calculation of the condensation temperature in the presence of disorder and assume the Bose gas to be exposed to a random potential $V(\x)$ with average $\langle V(\x)\rangle = 0$ and delta correlations $\langle V(\x_1)V(\x_2)\rangle = R_0\delta(\x_1-\x_2)$. This model was studied, e.g., in Refs.\ \cite{HuaMen92} and \cite{LopVin02} to describe He4 in porous media. The parameter $R_0$ is a measure for the strength of the disorder. To evaluate the partition function of the Bose gas we apply the replica method and consider an ensemble of $N$ copies of the system. The partition function for this ensemble, averaged over the random potential, is given by \cite{Pel04,GraPel05}
\be
\ep{{\cal Z}^N}= \left\{\prod_{\alpha=1}^N\int {\cal D}\phi_\alpha^*\int {\cal D}\phi_\alpha \right\} e^{-S^{(N)}[\phi^*,\phi]/\hbar},
\ee
with
\bea\label{action2}
&&S^{(N)}[\phi^*,\phi] = \int_0^{\hbar\beta}d\tau\int d^Dx\sum_{\alpha=1}^N\bigg\{\phi_\alpha^*(\bfx,\tau)\nln
&&\!\!\!\!\!\left.\times\left[\hbar\pder{}{\tau}-\frac{\hbar^2}{2M}\mathbf{\Delta}-\mu\right]\phi_\alpha(\bfx,\tau)+\frac g2 \phi_\alpha^{*\,2}(\bfx,\tau)\phi_\alpha^2(\bfx,\tau) \right\} \nln
&& -\frac{R_0}{2\hbar}\int_0^{\hbar\beta}d\tau\int_0^{\hbar\beta}d\tau'
\int d^Dx\nln
&& \times \sum_{\alpha=1}^N\sum_{\alpha'=1}^N
\phi_\alpha^*(\bfx,\tau)\phi_\alpha(\bfx,\tau)
\phi_{\alpha '}^*(\bfx,\tau')\phi_{\alpha '}(\bfx,\tau').
\eea
Again making use of the first-order derivative expansion, the RG approach outlined above can be extended to deal with the additional terms in Eq.\ (\ref{action2}) introduced by the random potential. The resulting flow equations now also describe the renormalization of $R_0$. They read
 \bea
\label{flowM2}\frac{dM}{dl}&=&2M+ \frac 1{2\pi^2}\left[-2\tG bN_{BE}+NbRN_{BE}\right.\nln
&& \left. +\frac R{1/2-M}\right],\\
\label{flowtG2}\frac{d\tG}{dl}&=& \tG+\frac 1{2\pi^2}\left\{-\tG^2b\left[4bN_{BE}(N_{BE}+1)\right. \right.\nln
&&\left.\left.+\frac{1+2N_{BE}}{2(1/2-M)}\right]+\tG R\frac 6{(1/2-M)^2}\right\},\\
\label{flow3}\frac{dR}{dl}&=& R+\frac 1{2\pi^2}\left[(NbR-4b\tG)RbN_{BE}(N_{BE}+1)\right.\nln
&& \left.+R^2\frac 4{(1/2-M)^2}\right],
\eea
with the scaled disorder parameter $R=\Lambda^3\beta_\Lambda^2R_0$. Note that the number $N$ of copies of the Bose gas now enters as a parameter into the flow equations. Within our one-loop approximation, the averaged partition function for the ensemble is given by
\be\label{avpart}
\ep{{\cal Z}^N} = \prod_{\k} \left(1-\exp\{-\beta[\veps_\k-\mu(|\k|;N)]\}\right)^{-N}
\ee
with $\veps_\k=\hbar^2\k^2/2m$ and $\mu(|\k|;N)=e^{-2l}M(l;N)/\beta_\Lambda$ the renormalized chemical potential with the trivial scaling removed. The parameters $k=|\k|$ and $l$ are related by $k=\Lambda e^{-l}$. The scaled free energy $W\equiv (\beta_\Lambda/\beta) \ep{\ln Z}$ of a single copy of the system is obtained through the replica limit
\bea
W&=&-\frac {\beta_\Lambda}\beta\lim_{N\to 0}\frac{\ep{{\cal Z}^N}-1}N\\
&=&\frac{1}{2\pi^2 b(0)}\int_0^\infty dl e^{-3l}
\ln{[1 - e^{- b(l) [1/2 - M(l;N=0)] } ] }\nonumber
\eea
In other words, the free energy is calculated by integrating the flow equations at $N=0$ [compare with Eq.\ (\ref{Whom})].

Analogous types of disorder, i.e., the so-called random exchange or random temperature defects, have been thoroughly examined in the context of classical spin systems \cite{HarLub74,Lub75,GriLut76,Khm76,Her85}. In the Ginzburg-Landau model, these defects are described by adding a random term $\delta r(\x) S^2(\x)$ to the Hamiltonian, where $S(\x)$ is the spin distribution and $\delta r(\x)$ is delta correlated just like the potentials $V(\x)$ above \cite{Her85}. It is therefore not surprising that in the high-temperature limit $b(l)\to 0$ the RG equations (\ref{flowM2})--(\ref{flow3}) reduce to the flow equations for the classical system (compare with Eqs.\ (2.9) of Ref.\  \cite{Her85}). 
In particular, the fixed-point structure of Eqs.\ (\ref{flowM2})--(\ref{flow3}) is the same as for the spin system.
For $N=0$, these equations possess three nontrivial asymptotic fixed points:\\
1. $(M^*_1,\tG^*_1,R^*_1)=(1/12, 5\pi^2/72, 0)$ with eigenvalues $((3 + \sqrt{249})/10,(3 - \sqrt{249})/10,\frac{1}{5})$. This is the Wilson-Fisher fixed point well known from the ordered system. However, due to the presence of disorder it becomes destabilized and no longer describes a true phase transition. This can be seen from the fact that two eigenvalues are positive.\\
2. $(M^*_2,\tG^*_2,R^*_2)=(3/22, 16\pi^2/121, 8\pi^2/121)$ with eigenvalues $( (1 + \sqrt{177})/8,(1 - \sqrt{177})/8,-\frac{1}{2}  )$. This so-called random fixed point \cite{HarLub74,Lub75} is associated with condensation in the presence of disorder.\\
3. $(M^*_3,\tG^*_3,R^*_3)=(1/18, 0, -2\pi^2/9)$. This fixed point does not have a physical meaning: it has $\tG^*_3=0$, but there is no phase transition in the disordered interaction-free Bose gas.

\begin{figure}
\begin{center}
\includegraphics[width=7.5cm]{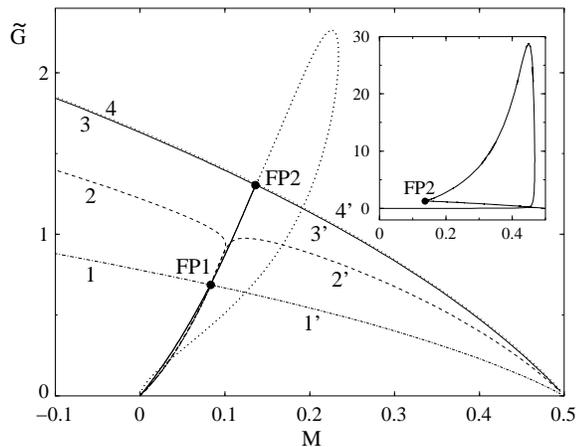}
\caption{Critical trajectories of flow equations (\protect\ref{flowM2})--(\protect\ref{flow3}) projected onto $M$-$\tilde G$ plane. The trajectories have $\ha = 4.9\times 10^{-4}$ and $\varrho=2.2\times 10^{-4}$ (dashed-dotted), $4.0\times 10^{-2}$ (dashed), $0.44$ (full), 1.78 (dotted), 20.0 (inset).
\label{fig1} }
\end{center}
\end{figure}
 
Figure \ref{fig1} illustrates the influence of disorder on the flow described by Eqs.\ (\ref{flowM2})--(\ref{flow3}), and shows the crossover between the fixed points (FPs) 1 and 2. The figure displays critical trajectories at fixed $\ha = a/\lambda_T = 4.9\times 10^{-4}$ and increasing $\varrho = (m\beta/4\pi\hbar^2) ( R_0/a) = b(0) R(0)/4\pi a\Lambda$. The parameter $\varrho$ is a convenient dimensionless measure of the strength of disorder relative to the particle interactions. For each value of $\varrho$, two very close numerical approximations $i$ and $i'$, $i=1,\dots,4$, to the actual critical trajectory are shown that eventually depart to the left and the right, respectively.

For $\varrho\ll 1$ (trajectory set 1 in Fig.\ \ref{fig1}), the critical flow is still essentially dominated by FP 1. Although this fixed point is no longer asymptotically attractive, the influence of disorder is so weak that we were not able to numerically find trajectories that go beyond FP 1 and approach FP 2 (although we expect such trajectories to exist: if a critical trajectory remains in the vicinity of FP 1 long enough, the influence of disorder should eventually drive it away from FP1 and towards FP2). For increasing disorder (trajectory set 2 in Fig.\ \ref{fig1}), the critical trajectories go beyond FP 1, but we still cannot find trajectories reaching FP 2. This is only achieved for $\varrho\approx 1$ (see set 3 and 4). In the regime of $\varrho\gg 1$, the flow trajectories display ``interaction-free" behavior for a long integration time (see inset). Only when $M$ has almost reached 0.5, the influence of the interactions becomes noticeable and bends the trajectories towards FP 2. In this case $M(l)$, $\tG(l)$ and $R(l)$ become very large. Since our RG approach is based on expanding the Euclidean action in $\tG(l)$ and $R(l)$, it is probably no longer valid in this regime. We note that a very similar overall behavior of the flow is also obtained for larger values of $\ha$. From these results, we conclude that our RG calculation is applicable in the regime of $\varrho\lsim 1$.

\begin{figure}
\begin{center}
\includegraphics[width=7.5cm]{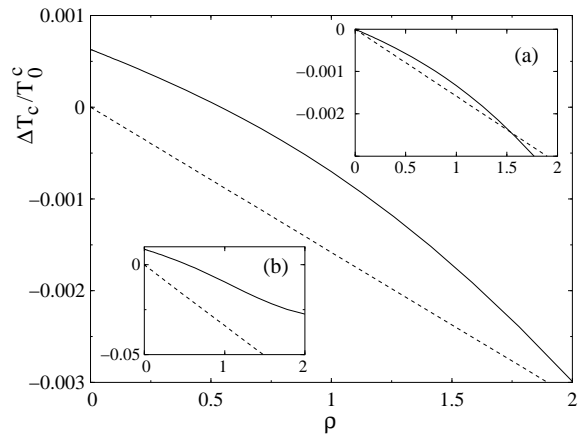}
\caption{Shift of critical temperature as function of $\varrho$ for fixed $\ha = 4.9\times 10^{-4}$. Full curve: RG result; dashed: Eq.\ (\ref{lopvin}). Insets: (a) RG result after removing offset at $\varrho=0$; (b) shift of $T_c$ for $\ha = 1.1\times 10^{-2}$.
\label{fig2} }
\end{center}
\end{figure}

The critical flow is characterized by the vanishing of the physical renormalized chemical potential [i.e., $M(l)e^{-2l}\to 0$ for $l\to\infty$]. Fixed point 2 can thus be associated with a crossover transition between the normal and a locally condensed phase \cite{LopVin02}. Using perturbation theory, in Ref.\ \cite{LopVin02} the corresponding shift in the condensation temperature due to disorder was estimated as
\be\label{lopvin}
\frac{T_c-T_c^{(0)}}{T_c^{(0)}} = -R_0 \frac{ T_c^{(0)} m^3}{6\pi^2n} = -\varrho \ha\frac{8\pi}{3\zeta(3/2)}\sqrt{\frac{T_c}{T_c^{(0)}}}
\ee
with $n$ the Bose gas density and $T_c^{(0)}$ the condensation temperature of the ideal gas. In the perturbative calculation, effects of particle interactions are neglected [note that the product $\varrho \ha$ on the right-hand side of Eq.\ (\ref{lopvin}) is independent of $a$]. However, the study of the fixed points and the critical flow in the RG approach suggest an essential role of the particle interactions even in the presence of disorder. It would thus be more satisfactory to calculate the shift within a nonperturbative approach that takes the interactions fully into account. With our RG method, such a calculation can be performed in an efficient manner.

In Fig.\ \ref{fig2}, we compare the RG calculation of $T_c$, which proceeds analogously to the case without disorder, to the perturbative result (\ref{lopvin}). The relative shift $\Delta T_c/T_c^0$ is shown at fixed $\ha$ as a function of the strength of disorder $\varrho$. We focus on the regime $\varrho\lsim 1$ where the RG approach is applicable. For $\varrho\to 0$, the RG shift smoothly tends to the (finite) value obtained in the calculation for the ordered interacting gas. This indicates that, in contrast to the perturbative calculation, our approach is able to simultaneously deal with the concurrent effects of interactions and disorder.

As we show in inset (a) of Fig.\ \ref{fig2}, if the $T_c$ shift at $\varrho=0$ is subtracted from the RG result, both curves are in relatively good agreement with each other. In particular, we find that for small, fixed $\ha$, the shift of $T_c$ in the RG behaves as $d_1^{{\rm (RG)}} \varrho+{\cal O}(\varrho^2)$. The numerical results suggest that the slope $d_1^{{\rm (RG)}}$ depends on the scattering length $\ha$, and for small $\ha$ the ratio to the corresponding value $d_1^{{\rm (pert)}}$ of the perturbative result (\ref{lopvin}) can be approximated as
\bdm
\frac{d_1^{{\rm (RG)}}}{d_1^{{\rm (pert)}}} \approx 0.7 -90\ha.
\edm  We thus find that in the RG calculation, $T_c$ decreases slower with $\varrho$ than predicted by perturbation theory. For increasing $\ha$, the discrepancy becomes larger, as can be seen from inset (b) in Fig.\ \ref{fig2}.


In summary, we have applied the momentum-shell RG method to examine two nontrivial questions regarding the condensation temperature of interacting Bose gases. First of all, we have calculated the shift in $T_c$ (and the critical chemical potential) for the homogeneous Bose gas without disorder up to second order in the scattering length. Secondly, we have presented a nonperturbative calculation of the condensation temperature of disordered Bose gases thereby taking the influence of interactions properly into account. Our results show that the RG method is an efficient and versatile tool to study critical properties of interacting Bose gases and presents an alternative to the approaches used so far.

Stimulating discussions with B.\ Kastening, A.\ Pelster, G.\ Metikas, and G.\ Alber are gratefully acknowledged.

\section*{APPENDIX}

In the following, the qualitative behavior of the critical chemical potential $v = (\beta\mu)_c$ in the RG description of the homogeneous Bose gas is worked out. It appears quite difficult to do this directly from Eqs.\ (\ref{flow1}) and (\ref{flowtG}). However, these equations can be simplified in a way which, on the one hand, does not modify the essential behavior of $v$ but, on the other hand, allows to investigate the dependence on $\ha\equiv a/\lambda_T$ analytically. In a first step, we use that typically $M\ll E_>$ with $E_>=1/2$ the scaled cutoff energy, and neglect the $M$ dependence of the Bose-Einstein distribution $N_{BE}$ and of the denominator $2(E_>-M)$ which appears in Eq.\ (\ref{flowtG}). Furthermore, we disregard the nontrivial renormalization of $\tilde G$ at high energies due to two-body scattering and thus drop the term $-\tilde G^2 b/2E_>$ from Eq.\ (\ref{flowtG}). The relation between the initial condition for $\tilde G(l)$ and the scattering length $\ha$ is then given by $\tilde G(0) = 8\pi^2\ha /\sqrt{2\pi b(0)}$. The flow equations now read
\bea
\label{flow11}\frac{dM}{dl}&=&2M- 2\frac{\tG}{2\pi^2}bN_{BE},\\
\label{flowtG1}\frac{d\tG}{dl}&=& \tG-\frac{\tG^2b}{2\pi^2}\left[4bN_{BE}(N_{BE}+1)  +\frac{N_{BE}}{E_>}\right]
\eea
with $N_{BE}=[\exp(b(l)E_>)-1]^{-1}$. In a second step, Eq.\ (\ref{flowtG1}) is simplified further by setting 
$$b(l)N_{BE}=\left\{ \begin{array}{cc} 
0 & \quad b(l)\ge 1, \\
1/E_> & \quad b(l)< 1.
\end{array}\right. $$ This modification, which has only a small effect on the flow of $\tilde G$, turns  Eq.\ (\ref{flowtG1}) into
\be\label{flowG3}
\frac{d\tG}{dl}=\left\{ \begin{array}{cc}
\tG & \quad l\le \ln \sqrt{b(0)}, \\
\tG -q \tG^2 & \quad l > \ln\sqrt{b(0)}
\end{array}\right.
\ee
where $q=20/2\pi^2$. The solution to this equation is given by
\be\label{solG}
\tG(l) =\left\{ \begin{array}{cc}
\tG(0)e^l & 
l\le \ln \sqrt{b(0)}, \\[0.1cm]
{\displaystyle \frac{\tG(0) e^l}{1+q \tG(0) [e^l-\sqrt{b(0)}]} } & 
l > \ln\sqrt{b(0)}.
\end{array}\right.
\ee
The critical initial condition $M_c(0)$ for the chemical potential is determined by the condition $e^{-2l}M(l\to\infty) = 0$. From Eq.\ (\ref{flow11}) thus follows that
\bea
M_c(0) &=&\frac 1{2\pi^2} \int_0^\infty 2\,\tG(l) b(l)N_{BE} e^{-2l} dl \non
&=&\frac 1{2\pi^2}\int_0^\infty  2\,\tG(0) e^l b(l)N_{BE} e^{-2l} dl\non
&& + \frac 1{2\pi^2}\int_{\ln \sqrt{b(0)}}^\infty 2\left(\frac{\tG(0) e^l}{1+q \tG(0) (e^l-\sqrt{b(0)})}\right. \non
&& \hspace*{0.8cm} - \tG(0)e^l \bigg)b(l)N_{BE} e^{-2l} dl\non
&\equiv& I_1 +I_2.
\eea
The first integral $I_1$ equals $b(0)^{-1} 4\zeta(3/2) \ha$, whereas $I_2$ can be rewritten as
\be\label{expansion}
I_2= \frac{q\tG(0)^2}{2\pi^2} \int_0^1 dx \frac{\sqrt x -1}{1-q\tG(0)\sqrt{b(0)}(1-1/\sqrt x)} \frac 1 {e^{xE_>}-1}. 
\ee
Expanding the second factor in the integrand as $1/(e^{xE_>}-1) = 1/xE_> -1/2 +\cdots$, one sees that expression (\ref{expansion}) can be written as
\be
I_2= \frac 1{b(0)} \left[\frac{640}\pi \ha^2\ln \ha + b_2'' \ha^2 +O(\ha^3)\right].
\ee
It is interesting to note that, similarly to the effective-field theory calculation of Ref.\ \cite{ArnTom01}, only the coefficient of $\ha^2\ln \ha$ can be determined by the perturbation expansion. All other terms apparently receive contributions from all expansion orders and thus cannot be calculated in this way. Putting together the above results and using numerical data to obtain the coefficient of the $\ha^2$ term, we finally find
\be
v=(\beta\mu)_c = 4\zeta(3/2) \ha + \frac{640}\pi \ha^2\ln \ha + 908 \ha^2 +O(\ha^3)
\ee
in the simplified approach studied here. It should be remarked that even in this simple model, the term linear in $\ha$ has the correct prefactor, whereas the coefficients of the two second-order terms differ by only a factor of about 2 from the values given in ref.\ \cite{ArnTom01} and the results of our original RG equations. We also note that from the numerical simulation of Eqs.\ (\ref{flow11}) and (\ref{flowG3}) we can accurately reproduce the coefficient $640/\pi$ of the $\ha^2\ln \ha$ term, thus confirming the accuracy of the fitting procedure described in the main text.

\end{document}